# The Metal–Silicate Partitioning of Carbon During Earth's Accretion and its Distribution in the Early Solar System

I. Blanchard[1#*], D. C. Rubie[1], E. S. Jennings[2], I. A. Franchi[3], X. Zhao[3], S. Petitgirard[1], N. Miyajima[1], S. A. Jacobson[4], A. Morbidelli[5]

[1] Bayerisches Geoinstitut, Universität Bayreuth, 95440 Bayreuth, Germany

[2] Department of Earth and Planetary Sciences, Birkbeck, University of London, Malet Street, London WC1E 7HX, UK

[3] School of Physical Sciences, Open University, Milton Keynes MK7 6AA, UK

[4] Department of Earth and Environmental Sciences, Michigan State University, East Lansing, MI 48824, USA

[5]Laboratoire Lagrange, Université Côte d'Azur, CNRS, Observatoire de la Côte d'Azur, 06304 Nice, France

[#]Now at Institut für Geowissenschaften, Universität Potsdam, 14476 Potsdam, Germany

*Corresponding author: blanchard@uni-potsdam.de

**Abstract**

Carbon is an essential element for the existence and evolution of life on Earth. Its abundance in Earth's crust and mantle (the Bulk Silicate Earth, BSE) is surprisingly high given that carbon is strongly siderophile (metal-loving) at low pressures and temperatures, and hence should have segregated almost completely into Earth's core during accretion. Estimates of the concentration of carbon in the BSE lie in the range 100-260 ppm and are much higher than expected based on simple models of core–mantle differentiation. Here we show through experiments at the putative conditions of Earth's core formation (49–71 GPa and 3600–4000 K) that carbon is significantly less siderophile at these conditions than at the low pressures (≤ 13 GPa) and temperatures (≤ 2500 K) of previous large volume press studies, but at least an order of magnitude more siderophile than proposed recently based on an experimental approach that is similar to ours. Using our new data along with previously published results, we derive a new parameterization of the pressure–temperature dependence of the metal–silicate partitioning of carbon. We apply this parameterization in a model that combines planet formation and core-mantle differentiation that is based on astrophysical N-body accretion simulations. Because differentiated planetesimals were almost completely depleted in carbon due to sublimation at high temperatures, almost all carbon in the BSE was added by the accretion of fully-oxidized carbonaceous chondrite material from the outer solar system. Carbon is added to the mantle continuously throughout accretion and its concentration reaches values within the BSE range (e.g. 140±40 ppm) at the end of accretion. The corresponding final core and bulk Earth carbon concentrations are 1270±300 ppm and 495±125 ppm respectively.

## 1. Introduction

Accretion and differentiation of the Earth occurred within the first 200 million years of the beginning of the Solar System and established the compositions of Earth's mantle and core. Earth's accretion involved a series of giant impacts, the energy of which resulted in one or more magma oceans with depths exceeding 1000 km (e.g. Rubie et al., 2015). The extent to which the volatile element contents of terrestrial planets were set by the processes of accretion and core formation, volatile loss, and the addition of a late veneer following core formation, are controversial (Albarède, 2009; Li et al., 2021; Rubie et al., 2015; Siebert et al., 2018; Wood et al., 2010). Various scenarios have been proposed according to which volatile elements were accreted to Earth either during the main phase of accretion (e.g. Suer et al., 2017), or during late accretion after core formation was complete (e.g. Albarède, 2009). By modelling core formation using experimental high pressure ($P$) and high temperature ($T$) data on the metal–silicate partitioning of elements that are both siderophile and volatile, it should be possible to discriminate between the two scenarios.

Carbon is both a volatile and a siderophile element, and as such has necessarily been affected by both core formation and volatilization processes during the accretion of Earth and its precursor bodies (e.g. Dasgupta et al., 2013; Li et al., 2021). Today, carbon is intimately linked to several key processes on Earth, such as the evolution of the atmosphere, the presence of life at the surface and the formation of diamonds in the mantle. Most estimates of the concentration of carbon in the BSE lie in the range 100–180 ppm (Hirschmann, 2018; McDonough, 2014; McDonough and Sun, 1995) although Marty et al. (2020) have proposed the possibility of a slightly higher concentration (112-260 ppm). At low $P$ and $T$, carbon behaves as a strongly siderophile element, with a partition coefficient well above 100 (e.g. Dasgupta et al., 2013), so that effectively all carbon should have partitioned into the core during its formation. This would leave the mantle strongly depleted in C, which is inconsistent with BSE estimates. Several explanations have been proposed for the high BSE C concentration, including crystallization of C-rich phases (e.g. diamond) from a C-saturated magma ocean (Armstrong et al., 2015; Kuwahara et al., 2021); imperfect metal–silicate equilibration during core formation (Hirschmann, 2012); accretion of a "late veneer" after the end of core formation (Dasgupta et al., 2013); carbon outgassing from the core (Dasgupta et al., 2013); and carbon becoming much less siderophile at high $P$–$T$ (Fischer et al., 2020).

Understanding the distribution of carbon in the Earth has been the subject of numerous experimental studies of metal–silicate partitioning. In the majority of cases, experiments have been performed using piston-cylinder or multi-anvil apparatus with the aim of determining the effects of composition (of both metal and silicate) on the metal–silicate partition coefficient $D_C$ (see Eq. 2) for carbon during core formation. The degree of silicate melt depolymerization, typically described by the low-pressure nbo/t parameter (number of non-bridging oxygen per tetrahedra), has been shown to affect $D_C$. As melt depolymerization and nbo/t increase, $D_C$ decreases (Chi et al., 2014; Dasgupta et al., 2013; Fichtner et al., 2021; Kuwahara et al., 2019). The presence of light elements in the metallic phase, such as sulfur, silicon, and nitrogen, also affects $D_C$ (Chi et al., 2014; Dalou et al., 2017; Grewal et al., 2020, 2019; Li et al., 2016a, 2015a). However, using piston-cylinder and multi-anvil apparatus, the effects of $P$–$T$ have not been fully investigated. Using such apparatus, the highest $P$–$T$ conditions were reached by Malavergne et al. (2019) (15 GPa and 2573 K), but these are still significantly lower than the $P$–$T$ conditions present during much of Earth's core formation (at least 40–60 GPa and 3500–4000 K, Siebert et al., 2012; Rubie et al., 2015). Recently, Fischer et al. (2020) performed diamond anvil cell experiments (LH-DAC) to determine metal–silicate partition coefficients for carbon at 37–59 GPa and 4200–5200 K. They concluded that C becomes much less siderophile as $P$ and $T$ increase. Measuring carbon in the quenched silicate melt of LH-DAC experiments is not straightforward, and Fischer et al. (2020) used both an electron microprobe (EPMA) and NanoSIMS. However, they found an order of magnitude difference between C concentrations in quenched silicate liquid measured by these two techniques. Consequently, there is significant uncertainty in their experimental results.

Here we present the results of LH-DAC experiments that determine the metal–silicate partition coefficient of carbon at conditions that are comparable to those of metal–silicate equilibration and segregation during core formation. We use the results in a state-of-the-art model that combines planetary accretion and core–mantle differentiation (Rubie et al., 2015), and show that the present BSE carbon concentration is the direct consequence of these combined processes. We conclude that carbon was added to Earth throughout accretion, a result that is contrary to the late veneer hypothesis.

## 2. Methods

### 2.1 Starting materials

Our starting silicate material consisted of a glass with the composition of a mid-ocean ridge basalt that was synthetized by mixing oxide and carbonate compounds in stoichiometric proportions, decarbonating this mixture overnight at 900°C, then adding FeO. Powders were ground under acetone to ensure a homogeneous composition. We created pellets that were subsequently fused in an argon flux for several seconds at about 1400°C using a levitation furnace device in Orléans, France (Auzende et al., 2011). The chemical composition of the glass was checked by EPMA ("Basalt" in Table S1) and no compositional variation was detectable. The resulting glass sphere was subsequently polished down to a thickness of 20 μm and machined in Institut de Physique du Globe de Paris, France, to obtain small disks of 80 μm diameter (Blanchard et al., 2017). The choice of a basaltic starting composition was made in order to synthetize quenchable glass standards, suitable for FTIR and NanoSIMS measurements. Although a peridotite composition would have been more applicable to core formation, peridotite melt cannot readily be quenched to a glass or extremely fine-grained crystalline material.

The metal starting material was synthesized from Fe plus 5 wt.% C. Carbon is a prominent contaminant in the laboratory, due to the use of ethanol and acetone, various types of glue, and ambient contamination. It has also been suggested that diamonds used as pressure transmitting anvils in DAC experiments can be the source of carbon that diffuses to the sample (Fischer et al., 2015). To better understand the origin of carbon, we used only $^{13}$C in our experimental samples in order to assess the extent of carbon contamination from external sources. We mixed 5 wt.% $^{13}$C powder (97% purity from Cambridge Isotope Laboratories, Inc.) with 95 wt.% Fe. We melted this alloy at 2 GPa and 1873 K in an MgO capsule for 10 minutes using a piston-cylinder apparatus. The quenched $^{13}$C-doped metal was then crushed and used as the metallic portion of the sample. The composition was checked by EPMA (Table S1, "Metal") using an Fe$_3$C standard (see section 2.5).

## 2.2 LH-DAC experiments

We performed four LH-DAC experiments at 49–71 GPa and 3600–4000 K. We used diamonds with 250 μm culets and rhenium gaskets. The gaskets were pre-indented to obtain a thickness of 40-50 μm, and were subsequently laser-drilled to produce a sample chamber ~90 μm in diameter. We loaded two silicate disks sandwiching a flake of Fe-$^{13}$C alloy into this chamber (Blanchard et al., 2017). Samples were compressed to the target pressure and then laser-heated at the desired temperature using a doubled-sided laser-heating system at BGI. Pressure was determined using the Raman shift of the diamonds (Akahama and Kawamura, 2004), which also allowed us to avoid using a ruby chip in the experiments. High temperatures were generated by two fiber continuous-wave (CW) YAG lasers (SPi©) with wavelengths of 1064 nm and delivering 100 watts each. Temperature was measured simultaneously and continuously on both sides of the diamond cell sample, using a spectro-radiometric technique (Benedetti and Loubeyre, 2004). Temperature was increased above the liquidus temperature of the sample, held for a few tens of seconds and then the lasers were switched off to promote quenching. The samples quenched extremely quickly due to the high thermal conductivity of the diamond anvils and the samples were subsequently decompressed slowly over several hours. We estimated that the temperature uncertainty was ±300 K (Boehler, 2000). The estimate of the final pressure was corrected for thermal pressure following $\Delta P$ = 2.7 MPa/K (Andrault et al., 2011; Fiquet et al., 2010). Additional details of the laser heating procedure are presented in the supplementary material.

**2.3 Sample recovery**

We used the Focused Ion Beam facility (FIB, dual beam Scios FEI©) at BGI to recover small lamellae of 3–5 μm thickness, about 40 μm in length and 20 μm wide, from the samples using a $Ga^+$ ion beam (Blanchard et al., 2017; Siebert et al., 2012). In the samples, we observed distinct quenched metal and silicate phases that had both been molten at high P–T (Fig. 1a). As observed in previous LH-DAC studies, the silicate melt quenched to a phase that appears homogeneous at the resolution of FEG-SEM imaging. The quenched silicate is often assumed to be a glass (e.g. Fischer et al., 2020) but is shown here to consist mostly of crystalline bridgmanite and amorphous calcium perovskite with a grain size of a few tens of nanometers (section 3.1).

## 2.4 Fabrication of NanoSIMS standards

We synthesized $^{13}$C-doped glass standards following a method developed by Yoshioka et al. (2015). We first mixed oxides in basaltic proportions and made a carbon-free glass by quenching after melting at 1600°C for two hours. This glass was analyzed by EPMA for major elements, and by FTIR to confirm that it was carbon-free ("B1" in Table S1). Subsequently, we performed piston-cylinder experiments to incorporate $^{13}$C into the basaltic glass. The source of $^{13}$C was oxalic acid enriched in $^{13}$C (99% purity, Cambridge Isotope Laboratories, Inc.), that was loaded along with the crushed basaltic glass in a Pt capsule and pressurized to 2 GPa and heated at 1600°C for 10 minutes. Using this procedure, we created two glass standards (B1145 and B1147) containing 785 ppm and 1263 ppm of carbon respectively, as measured by FTIR. For the FTIR measurements, we used an extinction coefficient (Fine and Stolper, 1986) of 69500 L mol$^{-1}$ cm$^{-1}$. We report the FTIR measurements along with EPMA analyses in Table S1 for the carbon-free (B1) and the two carbon-doped glasses (B1145 and B1147).

## 2.5 EPMA analysis

The details of the analytical procedure are described in the supplementary material. We report in Table 1 the full analyses of the quenched silicate and metal liquids in our samples. Some analytical totals are low, as often observed in similar studies (e.g. Blanchard et al., 2017; Suer et al., 2017), but these low totals are not explained by the thin samples, because a 3 μm thick lamella is thick enough to not lose electrons (Jennings et al., 2019). We did not attempt to measure the concentration of C in the quenched silicate melt of our samples by EPMA for several reasons: first there are no appropriate standards, and second the concentrations of C present in our samples are too low compared to the background level to be analysable by EPMA (see also supplementary material).

**2.6 NanoSIMS analysis**

The concentrations of carbon in the silicate melt were expected to be extremely low, due to its high siderophility. LH-DAC experiments produce very small samples, with silicate regions that are only a few micrometers across. Therefore, NanoSIMS was used to achieve analytical precision and spatial resolution and to quantitatively analyze the low carbon concentrations in the quenched silicate melt.

NanoSIMS measurements are sensitive to matrix effects, so we took particular care in synthesizing and using relevant standards (see Section 2.4). In addition to the two synthesized standards, we used two standards of natural rock samples containing known concentrations of carbon: a basaltic glass from the East Pacific Rise (ALV 981-R23) containing 405 ppm $CO_2$ and a piece of glass from the D'Orbigny meteorite containing 40 ppm C (see Table S1). During NanoSIMS sessions, we analyzed the standards each day to check for consistency and to confirm the reproducibility of the measurements.

Our samples were mounted on TEM Cu-grids. To perform NanoSIMS measurements, the grids were mounted on electrically-conductive tape as this provides a more uniform electric field over the sample compared with using a standard Cameca TEM mount. NanoSIMS analyses were performed using the CAMECA NanoSIMS 50L at the Open University (Milton Keynes, UK). Prior to analysis, an area of each sample was pre-sputtered using a focused primary beam of 16 keV $Cs^+$ ions with a probe current of 100 pA to remove surface contamination. The sizes of the pre-sputtered areas varied from 5×5 to 7×7 $\mu m^2$. Analyses were then carried out in spot mode by rastering a 50 pA Cs$^+$ beam onto inner 3×3 to 5×5 μm$^2$ areas, with secondary ions of $^{12}$C$^-$, $^{13}$C$^-$, $^{30}$Si$^-$, $^{24}$Mg$^{16}$O$^-$, $^{27}$Al$^{16}$O$^-$ and $^{56}$Fe$^{16}$O$^-$ collected in electron multipliers simultaneously. Only the data from the inner 60% of the areas were collected using electron gating to avoid carbon contamination from the surrounding areas. Spot analysis, rastering at high speed without the recording of spatial information within the analysis area, consisted of 200 measurements, with a total analysis time of ~2 minutes. The data were processed with L'Image software (Larry Nittler, Carnegie Institute Washington D.C.), correcting for detector deadtime and stage drift, after which we extracted from the region of interest (ROI) the ratios $^{13}$C/$^{30}$Si and $^{12}$C/$^{30}$Si. We derived a precise calibration line for the C/Si ratio using our standards, which was used to infer the concentrations of carbon (both $^{12}$C and $^{13}$C, see Fig. 3a,b) present in each of our samples.

We assess in the supplementary material the different sources of uncertainties that can arise from NanoSIMS measurements, and we used the calibration curves (see Fig. 3a, b) to estimate the uncertainties on each $^{13}$C and $^{12}$C measurements reported in Table 1.

Analysing our samples and standards by NanoSIMS revealed that the carbon present in both cases was not only $^{13}$C but also $^{12}$C, with a lower $^{13}$C/$^{12}$C ratio in the quenched silicate melt than in the metal (this ratio ranges from 2.8-6.9 for metal and 0.37-1.7 for silicate, see Fig. 3c). The presence of significant concentrations of $^{12}$C was unexpected, since we carefully prepared the samples using only $^{13}$C, but demonstrates that carbon contamination is probably ubiquitous in all high-temperature DAC experiments. The silicate starting powders were heated overnight at 900 °C and then converted to glass using an aerodynamic levitation system at temperatures of about 1400 °C. Thus, despite several hours at elevated temperatures, it seems extremely hard to eliminate the presence of environmental carbon. We discuss in the supplementary material the attainment of the chemical equilibrium. We also measured unmelted silicate in samples BAS C 39 and BAS C 42 at the NanoSIMS: these contained 386 ppm (378 ppm of $^{12}$C and 8 ppm of $^{13}$C) and 347 ppm (190 ppm of $^{12}$C and 157 ppm of $^{13}$C) of carbon respectively, indicating the presence of significant $^{12}$C in the starting material. This is also supported by the presence of substantial amount of $^{12}$C in our standards B1145 and B1147 (see table S1) despite the use of only $^{13}$C during their synthesis. We therefore interpret this to be a consequence of carbon contamination originating from the silicate or metal starting material, rather than the diamond anvils (Fischer et al., 2015; Suer et al., 2017), but more work is necessary to clarify this.

**2.7 TEM analysis**

We performed transmission electron microscope (TEM) imaging and analyses on a FEI Titan G2 80-200 TEM equipped with an X-ray energy dispersive spectrometer (EDS) and electron energy-loss spectrometer (EELS) in order to observe and analyze the quenched metal and silicate melts. To do so, we further thinned one of our samples to a thickness of 60-80 nm using FIB. The FIB lamella was plasma cleaned prior to TEM analysis in order to remove surface hydrocarbon contamination. We determined Fe/C ratios in this sample and also perform EDS mapping of the metallic phase and imaging of the sample. The quantification of the EELS analysis of carbon followed the procedure previously described (Fischer et al., 2015; Miyajima et al., 2009), using experimentally-determined ratios of partial cross sections of C K and O K edges against the Fe L edge versus sample thickness, which were calibrated with synthetic $Fe_{0.94}O$ and $Fe_3C$ samples. Note that due to the inhomogeneity of the metallic region of the sample (Fig. 1b, c, d), the EELS technique cannot provide a measurement of the bulk carbon concentration in this region.

### 3. Results

### 3.1 Quench textures

In order to understand the chemistry of the metallic phase, we obtained TEM images of sample BAS C 39 (Fig. 1). Electron diffraction analyses show the presence of stoichiometric $Fe_7C_3$ inclusions in the quenched metal (Fig. 1b). To further investigate the quench textures (Fig. 1c), we performed scanning transmission electron microscopy (STEM)-energy dispersive X-ray spectroscopy (EDS) mapping to identify exsolved Si-O-rich inclusions in the metal (Fig. 1d) along with a C-rich domain that could be an exsolved diamond, as observed in previous similar experiments (Lord et al., 2009).

In the silicate phase of the thinned sample BAS C 39, a quench texture is observed. This texture is only visible by careful TEM sample preparation and imaging before amorphization of bridgmanite occurs, and is not observable by SEM, which may explain why such a texture has not been reported previously. The molten silicate quenched to a fine-grained crystalline phase (grain size of a few 10s of nanometres, see Fig. 2a) and not to a glass as has usually been assumed. The quenched phase was analysed by EDS and, based on its composition, is identified as Fe-rich bridgmanite. The amorphous zones between the quench crystals are caused by $Ga^+$-beam damage to the unstable bridgmanite crystals during the FIB procedure (Fig. 2a, see also Engelmann et al., 2003). Given the very small grain size of the quench crystals relative to the NanoSIMS spot size, we are confident that our measurements of C abundances are reliable.

We observe two distinct populations of metallic inclusions in the quenched silicate melt: (1) relatively large irregularly dispersed inclusions 100 to several 100 nm across, and (2) much smaller uniformly-dispersed inclusions a few tens of nm across (Fig. 2). Only the large inclusions are visible by SEM imaging, which may explain why this bimodal size distribution has not been described previously. The large metallic inclusions in the silicate have been described in previous studies (e.g. Blanchard et al., 2017), and have usually been considered to have exsolved from the silicate melt during quenching. Our STEM EDS measurements indicate that the large metal inclusions have compositions similar to that of the central main metallic region of the samples. We therefore interpret these large inclusions as having been present at high temperature during the experiment as entrained metal. Hence, we avoided these inclusions when analysing the quenched silicate by both NanoSIMS and EPMA.

In contrast, the small metal inclusions are unavoidable and are uniformly dispersed throughout the skeletal quench texture, and were included in our silicate melt measurements using NanoSIMS and EPMA. We interpret the small uniformly-dispersed metallic inclusions in the silicate as having formed during the quench, possibly due to bridgmanite crystallisation which caused their exsolution due to $Fe^{2+}$ disproportionation by the reaction $3Fe^{2+} \rightarrow 2Fe^{3+} + Fe^0$ (Frost et al., 2004).

### 3.2 Thermodynamics

We calculated the oxygen fugacity of our experiments relative to the iron-wüstite redox buffer (ΔIW), based on the assumption of ideal mixing of both phases (as justified by the high temperatures), with $X_{FeO}$ and $X_{Fe}$ being the molar fractions of FeO and Fe in the silicate and metal respectively:

$$\Delta \text{IW} = 2 \log \frac{X_{\text{FeO}}}{X_{\text{Fe}}}. \tag{1}$$

Oxygen fugacities lie in the range -0.9 to -1.5 log units relative to the *IW* buffer (Table 1). The concentrations of carbon ($^{12}$C+$^{13}$C) in the metal lie in the range 5 to 9.4 wt.% and in the quenched silicate melt 545 to 2800 ppm. The concentrations of carbon in the quenched silicate melt are much higher than concentrations observed in low *P–T* experiments which are typically 10–200 ppm (Armstrong et al., 2015; Chi et al., 2014; Dasgupta et al., 2013; Stanley et al., 2014). Thus, $D_c$ values lie in the range 23–157 (Table 1), and are 1-2 orders of magnitude lower than those determined in previous low-pressure studies.

In the following, we use the concentration of carbon in the silicate obtained from nanoSIMS analysis (the sum of $^{12}$C and $^{13}$C) and the concentration of carbon in the metal as measured with EPMA analysis to calculate $D_C$ of our experiments.

To understand the effects of *P–T* on $D_c$, we compiled a carefully-selected dataset from previous studies to supplement our new results (Fig. 4). Only some published experiments are directly relevant to core formation because prior studies have often focussed on the effects of silicate melt composition (Chi et al., 2014), $fO_2$ (Li et al., 2017; Malavergne et al., 2019), and the interaction of carbon with elements in the metal (e.g. Li et al., 2016b). The partitioning of carbon is potentially affected by silicate melt structure which is conventionally described by the low-pressure parameter nbo/t (Chi et al., 2014). Because of the use of a MORB as starting silicate, our samples lie in a nbo/t range of 0.56–1.34. We therefore only included partitioning data from the literature for which nbo/t lies in the range 0.5–1.5. We also only included data from previous low *P–T* studies with low concentrations (< 1 wt.%) of light elements (S, N, O, Si) in the metal because the interaction between carbon and other light elements may be significant (Li et al., 2015b) but is poorly constrained. As a consequence, we also excluded very low $fO_2$ results ($\Delta IW < -3$), as they are often associated with high concentrations of Si in the metal.

We have not included results obtained at similarly high $P$–$T$ conditions by LH-DAC (Fischer et al., 2020), who proposed very low $D_c$. Fischer et al. (2020) measured the concentration of carbon in quench silicate melt by both EPMA and NanoSIMS and the measurements differ by about an order of magnitude (<0.8 wt.% by EPMA and several wt.% by NanoSIMS). They state that the NanoSIMS data are likely the more accurate. Using NanoSIMS, they analysed large areas of the quenched silicate melt that included the large metallic blebs which they considered to have resulted from the quench, implying that they would have been dissolved in the silicate at high $P$–$T$. By contrast, we avoided including the large metal blebs when analysing the silicate because we interpret them to have been present in the silicate melt at high $P$–$T$ (section 3.1). These differing textural interpretations and analytical approaches are likely to be responsible for the significantly different measurements of C concentrations in the silicate melt. Notably, C is siderophile and thus resides predominantly in the metal blebs: including them in the silicate analysis therefore results in much higher C concentrations than when they are excluded. Including the metal blebs in their analysis also results in higher FeO content in the silicate melt of their experiments compared to ours. We conclude that these different interpretations of the large metal blebs are responsible for the fact that the derived $D_c$'s of Fischer et al. (2020) are lower than ours by about an order of magnitude. We compare our data and fitted models with those of Fischer et al. (2020) in the supplementary information (Figs. S3 and S4).

Our high $P$–$T$ samples contain significant concentrations of oxygen and silicon in the metal, as is typical for high temperature experiments (Blanchard et al., 2017; Siebert et al., 2012; Fischer et al., 2015). By combining the two datasets (low and high $P$–$T$), we thereby empirically include the interaction effects of these light elements in the regressions presented below. The concentrations of O and Si in liquid metal are generally low at low *P–T* (e.g. Frost et al., 2010) as is the case with the data that we have selected from previous C partitioning studies (Armstrong et al., 2015; Chi et al., 2014; Dasgupta et al., 2013; Stanley et al., 2014), whereas at high *P–T* such concentrations become high (Siebert et al., 2013), which is consistent with results of the present study. Interactions of carbon with oxygen and silicon in the metal of our samples might be at least partly responsible for the scatter of our data (Fig. 4). In Fig. 4, we present a comparison of our data together with the data selected from the literature as a function of *P–T*.

The metal–silicate partition coefficient of carbon ($D_C$) is calculated as:

$$D_\text{C} = \frac{X_\text{C}^\text{metal}}{X_\text{C}^\text{silicate}}, \qquad (2)$$

where *X* is the mole fraction of the element in the phase of interest. $D_C$ is a function of several parameters, including *P*, *T*, and oxygen fugacity.

To remove the effect of oxygen fugacity, we use the distribution coefficient $K_D$:

$$K_\text{D} = \frac{D_\text{C}}{(D_\text{Fe})^{n/2}}, \qquad (3)$$

where *n* is the valence of C when dissolved in silicate liquid. The valence of carbon at these conditions is uncertain, although a valence of 2+ is most likely (Armstrong et al., 2015; Grewal et al., 2020; Malavergne et al., 2019; Yoshioka et al., 2019). It has also been proposed that carbon can have a valence of 4+ (Shcheka et al., 2006), so we performed regressions using both valence states. Previous studies have suggested the presence of $CH_4$ in silicate melts (e.g. Li et al., 2015), but this has been proposed at very low $fO_2$ conditions and at low *P–T* (e.g. below

IW-2, Li et al., 2017) and has never been assessed at the conditions of our study. Hence, we do not consider CH$_4$ as a possible speciation in our samples.

Based on the aforementioned data selection, we performed a least-squared equal weight regression using 43 previously–published data (Armstrong et al., 2015; Chi et al., 2014; Dasgupta et al., 2013; Stanley et al., 2014) together with our new results to derive the following expressions with a valence of 2+ (Eq. 4) and 4+ (Eq. 5) :

$$\log K_D^{2+} = 0.3(\pm 0.2) + \frac{3822(\pm 860)}{T} \qquad (4)$$

$$\log K_D^{4+} = -1(\pm 0.5) + \frac{4842(\pm 920)}{T} + 31(\pm 19)\frac{P}{T}. \qquad (5)$$

When refining the pressure effect in the case of a 2+ valence, the resulting parameter is considerably smaller than its uncertainty so we obtained the above result assuming no pressure effect. This does not mean that there is no pressure effect, but rather that we cannot separate it from the temperature effect because the effects of $P$ and $T$ are strongly correlated. Equations 4 and 5 show that carbon becomes less siderophile with increasing depth along a magma ocean geotherm, and predict $K_D$ values reasonably well for both low and high $P$–$T$ experiments (Fig. 5). The uncertainties on the parameters represent 95% confidence limits. In Fig. 5 we compare the agreement between the experimentally determined $K_D$ and the $K_D$ values predicted by Equations 4 and 5.

## 4. The behaviour of carbon during Earth's accretion

There are currently two competing models for Earth's formation. (1) Classical accretion models in which Mars-size embryos and much smaller planetesimals collide in the protoplanetary disk and grow to form the Earth and other terrestrial planets within a time period of 100-200 million years (e.g. Raymond et al., 2009). (2) Pebble accretion by which Earth is predicted to have accreted most of its mass extremely rapidly in only a few million years (Schiller et al., 2018). In a simple model of pebble accretion, Earth accretes very rapidly, and grows maintaining a magma ocean in which new pebbles are incorporated, adding small amounts of metal. This metal sinks to the bottom of the magma ocean and equilibrates, before further diapirs bring it to the core. This scenario is close to a continuous core formation model discussed in the supplementary material, and demands very large and late addition of carbon. As argued in the supplementary material, there is no reason to expect this. Hence, here we use the "Grand Tack" model (a variant of the classic accretion simulations) that successfully reproduces the physical and chemical properties of Earth, Venus and Mars (O'Brien et al., 2014; Rubie et al., 2016, 2015; Walsh et al., 2011).

We have applied the parameterizations of Equations 4 and 5 in a model of combined accretion and core–mantle differentiation (Jennings et al., 2021; Rubie et al., 2016, 2015), with modifications as described below. This model is based on N-body accretion simulations that start with 80–220 Mars-size embryos (initially located at 0.7-3.0 AU) that are embedded in a protoplanetary disk consisting of a few thousand much smaller planetesimals that are initially distributed over a heliocentric distance of 0.7 to ~9.5 AU. There are no starting bodies between 3.0 and 4.5 AU because this region was cleared out by Jupiter and Saturn. In the simulations, planets grow from the starting embryos through collisions with other embryos and planetesimals. Each collision potentially involves a core formation event during which accreted metal equilibrates with silicate liquid in a magma ocean and then segregates to the proto-core. By modelling metal–silicate equilibration based on mass balance and element partitioning (Rubie et al., 2011), the evolving compositions of the mantles and cores of all accreting bodies are tracked throughout the accretion process. The approach requires that the bulk compositions of all starting bodies are defined. The relative concentrations of non-volatile elements are assumed to match those of CI chondrites but with a 22% enrichment of refractory element concentrations (Rubie et al. 2011). An oxidation gradient in the protoplanetary disk defines the oxygen content of starting bodies, which is a major compositional variable because of its effect on oxygen fugacity during metal–silicate equilibration. Compositional parameters are refined by a least squares fit of the mantle composition of an Earth-like planet to that of the bulk silicate Earth. Such fits show that embryos and planetesimals that formed within ~1 AU of the Sun are highly reduced but with increasing heliocentric distance are increasingly oxidized (Monteux et al., 2018; Rubie et al., 2015). A decrease in the core mass fraction of iron meteorite parent bodies with increasing heliocentric distance is consistent with this oxidation gradient (Table S2 in Morbidelli et al., in press). Planetesimals originating from beyond 4.5 AU include both bodies that underwent early core-mantle differentiation (iron meteorite parent bodies) and fully-oxidized bodies that contain no metal and have an $H_2O$ content of 20 wt.%. Including differentiated planetesimals from the outer solar system (CC region) and undifferentiated planetesimals from the inner solar system (NC region) is new because such bodies were previously not included in the model (Jennings et al., 2021; Rubie et al., 2015, 2016).

Defining the distribution of carbon in the early solar system is crucial for the current model. Based on meteoritic studies, there are distinct chemical and isotopic differences between parent bodies that formed in the inner and outer regions of the solar system – as represented by noncarbonaceous (NC) and carbonaceous (CC) meteorites respectively (e.g. Kruijer et al., 2017). The carbon content of fully-oxidized (metal-free) planetesimals from the CC reservoir (>4.5 AU), as represented by CI chondrites, is 2.2-3.5 wt% whereas concentrations from undifferentiated bodies from the NC reservoir, as represented by ordinary chondrites, is 0.04-0.53 wt% (Hirschmann et al., 2021). However, C concentrations in iron meteorites from both reservoirs are extremely low (0.0004-0.06 wt%) and indicate the early loss of most C from differentiated planetesimals by sublimation at high temperatures (Li et al., 2021). We have estimated the bulk concentrations of carbon in iron meteorite parent bodies using estimates of their core mass fractions (Table S2 in Morbidelli et al., in press) and core C concentrations (Table S2 in Hirschmann et al., 2021), with the assumption that mantle concentrations are effectively zero. The large difference between carbon concentrations in differentiated and undifferentiated planetesimals (Table 2) means that the fraction of differentiated planetesimals is a crucial parameter. For simplicity, we assume that this fraction is the same for both NC and CC regions. In addition, we assume that the C bulk concentration of differentiated embryos is identical to the NC differentiated planetesimal value. Enstatite chondrites (ECs) contain 3-6 wt.% carbon. However, despite their isotopic similarity to the Earth, it is unlikely that they were major building blocks of the Earth given that they have a different chemical composition. Notably, in addition to high Si concentrations, they are richer in volatiles, probably as a consequence of their late formation in the disk.

In the model, the metallic cores of differentiated bodies equilibrate with only a small fraction of the silicate mantle of the target body after an accretional collision event (Rubie et al., 2015). This fraction is quantified using a hydrodynamic model of the interaction of metal and silicate as an impactor's core sinks in the target body's magma ocean (Deguen et al., 2011; Rubie et al., 2015). When fully-oxidized bodies (from beyond 4.5 AU) are accreted there is no core-forming event because they contain no metal and, importantly, siderophile elements, including carbon, are mixed into the protoplanet's magma ocean/mantle. Currently there is no model that describes quantitatively the chemical consequences when undifferentiated metal-bearing planetesimals from the NC region are accreted. The small metal grains in such material will be more widely dispersed in the magma ocean and may equilibrate with a larger fraction of the mantle compared with predictions of the Deguen et al. (2011) model. Nevertheless, the results presented here are obtained using the Deguen et al. (2011) model (based on the mass fraction of metal in accreting undifferentiated bodies) because this enables an upper limit to the final mantle C concentration to be determined (see below, supplementary material and Fig. S6).

Here, our aim is to reproduce the BSE C concentration. Most estimates lie around 100 ppm, with Hirschmann (2018) providing a recent estimate of 140±40 ppm, which is close to the estimation of Marty et al., (2020) who determined a BSE C concentration of 112-260 ppm.

We used the "Grand Tack" accretion simulation "4:1-0.5-8" (Rubie et al., 2015) with the same model parameters as in the study of Rubie et al. (2016). In this simulation, the final giant impact on the Earth-like planet (at ~1 AU) occurs at 113 Myr and involves an impactor with a mass of $0.12M_e$. Following this event, late accretion adds an additional mass of 0.3% to Earth, which successfully reproduces highly siderophile element (HSE) concentrations in the mantle (Rubie et al., 2016).

The final mantle and bulk Earth carbon concentrations are shown in Fig. 6a as a function of the fraction of differentiated planetesimals. If undifferentiated CC bodies contain on average 3.35 wt.% C, a BSE C concentration of 140±40 ppm is achieved when the percentage of differentiated planetesimals is 45±15%; the corresponding bulk concentration is 500±100 ppm and the corresponding concentration in the core is 1270±300 ppm. If undifferentiated CC bodies contain on average 2.2 wt.% C, the same BSE concentration is achieved when the percentage of differentiated planetesimals is 25±15%; the corresponding bulk concentration is then 600±150 ppm. These bulk C concentrations are close to estimates for Earth of 530±210 ppm (Marty, 2012) and 730 ppm (McDonough, 2014). Note that the results of Fig. 6a are inconsistent with a BSE C concentration as high as 260 ppm, as proposed by Marty et al. (2020), because this would require that there were no differentiated CC planetesimals - on the contrary, iron meteorites show that differentiated CC planetesimals did indeed exist. Finally, if the fraction of mantle that equilibrates with the accreted metal particles in undifferentiated NC planetesimals is increased by a factor of 5 or 10 (for example) compared with the prediction of the Deguen et al. (2011) model, the final mantle C concentration is reduced by only ~20-30 ppm (Fig. S6). Our results show that C is not extracted efficiently from the mantle during core formation in spite of its siderophile behaviour, and the mantle concentration increases continuously during accretion (Fig. 6b); this is in strong contrast to result of a simple continuous core formation model (Supplementary Material, Fig. S5). Two critical features of our model explain this major difference. First, undifferentiated CC bodies from the outer solar system contain water ice, are fully oxidized and contain no metal. Thus, there is no core formation event when they are accreted and the delivered C remains in the magma ocean/mantle. Second, when differentiated metal-bearing bodies are accreted, the metal of the impactor only equilibrates with a fraction of the target's mantle, the value of which depends on the size of the impactor's core and the depth of the magma ocean (Deguen et al., 2011; Rubie et al., 2015). The fractions of the mantle that equilibrate with a batch of accreted metal in the present simulation are 0.16-2.5% for planetesimal impacts and 2.6-9.9% for embryo impacts. Consequently, core formation events are extremely inefficient at removing carbon from the bulk of the mantle and transferring it to the core. These two critical features are absent in "continuous core formation" models that are often applied in studies of Earth's differentiation. In such models it is generally assumed that all accreted material contains metal which equilibrates chemically with the entire Earth's mantle. Consequently, such models predict that essentially the entire BSE carbon budget has to be delivered during late accretion, after core formation has ended (see supplementary material), as proposed by Albarède (2009). We know of no astrophysical accretion simulations that would support this scenario which requires the accretion of carbon-poor material during core formation and carbon-rich material afterwards, during late accretion. The continuous delivery of carbon during Earth's accretion is also supported by previous work on volatiles and noble gases elements (Marty, 2012). In addition, the hypothesis that carbon and other volatiles were delivered to Earth primarily by the Moon-forming impactor (Grewal et al., 2019) is also not supported by our results.

There are obviously significant uncertainties on the parameters listed in Table 2. However, the C concentration in undifferentiated CC planetesimals is the main parameter that controls the final mantle concentration. The effect of decreasing this parameter from 3.35 to 2.2 wt.% is shown in Fig. 6a. If the value is reduced to zero, the final mantle C concentration decreases to ~10 ppm, which confirms that accreted undifferentiated CC planetesimals are the main source of BSE carbon.

Additional sources of uncertainty, including the effects of using the partitioning model of Fischer et al. (2020), are discussed in the supplementary material. Uncertainties also arise because the effects of atmosphere formation and evolution are not considered and our treatment of undifferentiated NC planetesimals only provides an upper limit to the final mantle C concentration (Fig. S6). When these are accounted for in future models, the estimated fraction of differentiated CC planetesimals is likely to decrease.

Our results have major implications for the accretion of water to Earth. With 20 wt.% $H_2O$ in fully-oxidized CC bodies, we obtain a plausible mantle water concentration of 570±220 ppm and a core hydrogen concentration of 38±10 ppm, assuming that the metal–silicate partition coefficient for hydrogen is infinitely large (Rubie et al., 2015). Tagawa et al. (2021) determined the metal–silicate partition coefficients of hydrogen at high *P* and *T* and, in contrast to our result, concluded that 0.3-0.6 wt.% hydrogen partitioned into the core during Earth's accretion. Their differentiation model ignores the likelihood that water was accreted to Earth primarily in a small quantity of fully-oxidized CC material that was not affected by core formation and consequently they obtained a core concentration >2 orders of magnitude larger than our estimate.

## 5. Conclusions

The results of LH-DAC experiments at high *P* and *T* show that carbon becomes significantly less siderophile at the conditions of Earth's core-mantle differentiation compared with low *P*–*T* determinations. Using the new partitioning results in a model that combines accretion and core-mantle differentiation, we successfully reproduce the estimated BSE carbon concentration (e.g. 140±40 ppm) and show that carbon was delivered to Earth's mantle throughout accretion. The hypothesised late delivery of most BSE carbon after the end of core formation is therefore not required. The accretion of fully-oxidized planetesimals from the outer solar system, beyond 4.5 AU, is responsible for adding almost the entire BSE C budget because, in the absence of metal, there are no associated core formation events. In contrast, the accretion of metal-bearing planetesimals adds very little C to the mantle as C is transferred to the core because of its siderophile behavior. In future models, the effects of atmosphere formation should be included in order to have a more complete view on the fate of carbon during Earth's accretion.

**Acknowledgments**

We are grateful to Louis Hennet who made the glass starting material, Raphael Njul for sample preparation, James Badro for the access to the disk cutting facility in IPGP, Vera Laurenz and Detlef Krauße for their help at the microprobe, and Dan Frost for discussions. Takahiro Yoshioka and Svyatoslav Shcheka are thanked for their help at the piston-cylinder. Takahiro Yoshioka and Hans Keppler are acknowledged for the FTIR measurements. IB and SP were funded by the German Science Foundation (DFG) Priority Program SPP1833 "Building a Habitable Earth" (Ru1323/10-1, PE 2334/1-1). IB has also received support from the European Union's Horizon 2020 research and innovation program to access the Open University NanoSIMS facility at Milton Keynes, UK. Europlanet 2020 RI has received funding from the European Union's Horizon 2020 research and innovation program under grant agreement No 654208. Support was also provided by the European Research Council (ERC) Advanced Grant "ACCRETE" (contract 290568). The Scios FIB and the Titan G2 TEM at BGI (Bayerisches Geoinstitut) were financed by DFG Grants INST 91/315-1 FUGG and INST 91/251-1 FUGG, respectively. We thank Bernard Marty, Marc Hirschmann and an anonymous reviewer for their helpful and constructive reviews of our manuscript, and Bill McKinnon for editorial handling.

# Tables

Table 1: Compositions of quenched silicate and metal liquids together with experimental conditions of each experiment. Major elements were analysed using EPMA without coating, and carbon in the silicate was analysed using NanoSIMS.

| Run # | BAS C 50 | BAS C 60 | BAS C 39 | BAS C 42 |
|---|---|---|---|---|
| *Silicate phase wt.%* | *N=4* | *N=5* | *N=8* | *N=6* |
| $SiO_2$ | 35.84 (0.38) | 39.97 (1.98) | 34.09 (0.83) | 39.66 (0.99) |
| MgO | 8.85 (0.20) | 8.60 (0.28) | 9.53 (0.51) | 9.02 (0.26) |
| $Al_2O_3$ | 19.46 (0.12) | 25.93 (0.50) | 16.34 (1.36) | 24.48 (0.64) |
| CaO | 4.02 (0.16) | 4.89 (0.12) | 5.67 (0.20) | 5.68 (0.31) |
| $Na_2O$ | 2.13 (0.07) | 2.22 (0.10) | 2.32 (0.15) | 2.53 (0.11) |
| FeO | 23.82 (1.40) | 18.60 (0.60) | 27.35 (2.55) | 13.71 (1.72) |
| $^{13}C$ (ppm)[a] | 133 (10) | 150 (10) | 1780 (80) | 171 (10) |
| $^{12}C$ (ppm)[a] | 178 (10) | 394 (80) | 1022 (200) | 100 (10) |
| Total | 94.12 (1.85) | 100.2 (1.96) | 95.30 (2.34) | 95.07 (1.55) |
| *metal wt.%* | *N=3* | *N=3* | *N=4* | *N=2* |
| Fe | *81.68 (1.25)* | *79.56 (0.9)* | *85.96 (0.58)* | *78.01 (0.78)* |
| Si | *0.46 (0.16)* | *5.56 (0.16)* | *0.33 (0.11)* | *2.65 (0.40)* |
| O | *3.79 (0.89)* | *7.56 (0.37)* | *2.85 (0.48)* | *3.42 (0.10)* |
| C | *5.90 (0.18)* | *4.99 (0.28)* | *9.44 (0.24)* | *5.35 (0.20)* |
| Total | *91.83 (0.41)* | *97.66 (0.38)* | *98.57 (0.29)* | *89.43 (0.48)* |
| P (GPa)[b] | 60 | 71 | 49 | 56 |
| T (K)[b] | 3800 | 4000 | 3600 | 3700 |
| ΔIW[c] | -1.06 | -1.2 | -0.89 | -1.51 |
| $D_C$[d] | 141 ±7 | 65 ±14 | 23 ±1 | 157 ± 6 |
| $K_D$[e] | 41.9 ±1 | 16.28 ±1.4 | 8.13 ±1 | 27.6 ±1.1 |
| nbo/t[f] | 0.95 | 0.58 | 1.34 | 0.56 |

[a]Uncertainties based on uncertainty of the slope of the calibration curve of the NanoSIMS (see supplementary text for more details).

[b]The uncertainties on temperature and pressure measurements are estimated to be about 300 K and 5 GPa respectively.

[c]ΔIW is the calculated $fO_2$ relative to the IW buffer

[d]$Dc$ is the metal–silicate partitioning of carbon (mol.% of carbon in the metal / mol.% carbon in the silicate) and was calculated on a single cation basis using the sum of $^{12}C$ and $^{13}C$

present in the silicate measured by nanoSIMS and the C concentration in the metal measured by EPMA.

[e]$K_D$ is the distribution coefficient where $K_\mathrm{D} = \frac{D_\mathrm{C}}{(D_\mathrm{Fe})^{n/2}}$ with a valence $n= 2+$, following the reaction $Fe^{metal} + CO^{silicate} = C^{metal} + FeO^{silicate}$

[f]nbo/t is the number of non-bridging oxygen atoms over tetrahedrally coordinated cations (Mysen et al., 1982).

Table 2: Estimates of bulk carbon concentrations of starting planetesimals and embryos used in the accretion/core-mantle differentiation model.

|  | Undifferentiated | Differentiated |
|---|---|---|
| Carbonaceous chondrite (CC) bodies (>4.5 AU) | 2.2-3.35 wt.%[1] | 0.002 wt.%[3] |
| Noncarbonaceous chondrite bodies (NC) (<3 AU) | 0.16 wt.%[2] | 0.004 wt.%[3] |

[1] CM and CI meteorite concentrations respectively
[2] Average of values list by Hirschmann et al. (2021, Table S1).
[3] Average of values calculated from seven iron meteorite parent body core concentrations (Hirschmann et al., 2021, Table S2, "liquid C" column) together with determinations of corresponding iron meteorite parent body core mass fractions (Morbidelli et al., in press., Table S2).

**Figures**

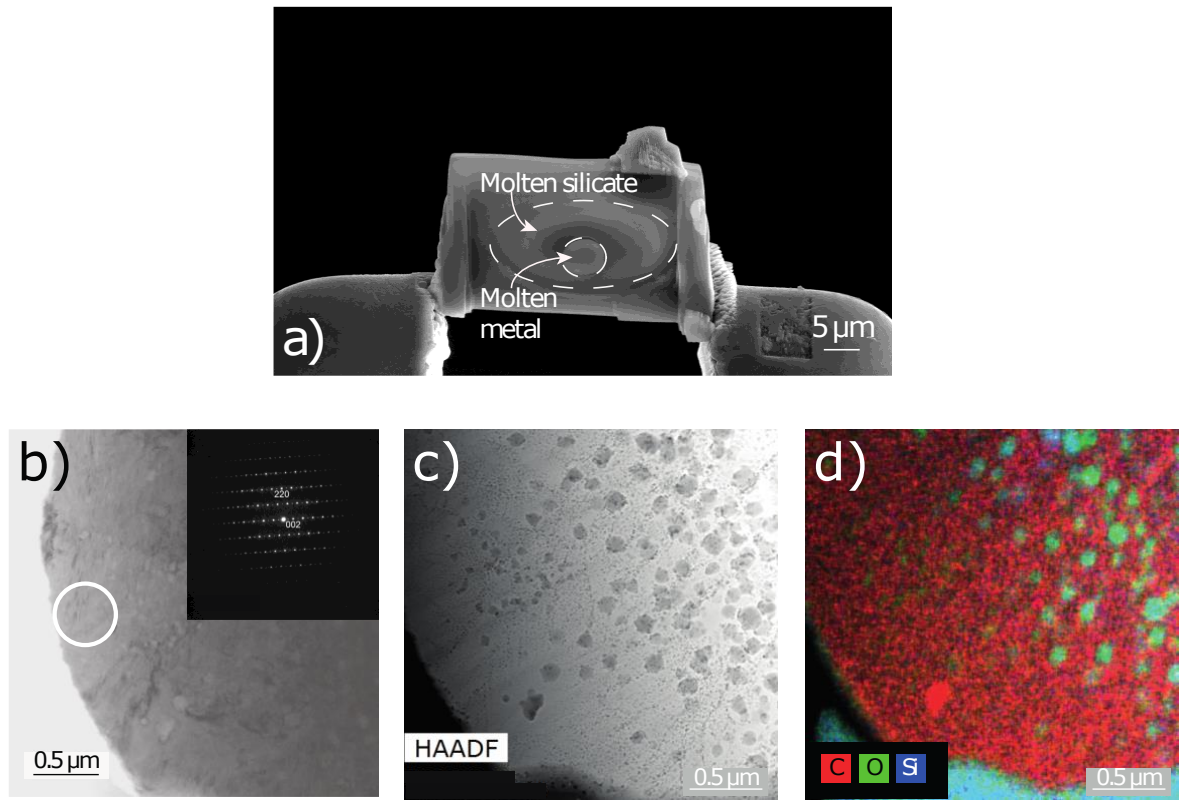

Figure 1: Images of sample BAS C 39 synthesized at 49 GPa and 3600 K. a) Secondary electron image in which two phases can be observed in the sample: the molten silicate and the molten metal that were equilibrated at high pressure and high temperature. b) Bright field TEM image of the metal: the white circle shows where the selected electron diffraction pattern (inset) that shows the presence of stochiometric $Fe_7C_3$ was taken (also confirmed with EELS). c) High-angle annular dark-field STEM image of the same metallic phase showing exsolutions (grey) that developed during quenching. d) STEM-EDS map of the same area which shows exsolved inclusions that are oxygen-rich (green) and silicon-rich (blue). The large bright red inclusion in d) is believed to be an exsolved diamond (Lord et al., 2009).

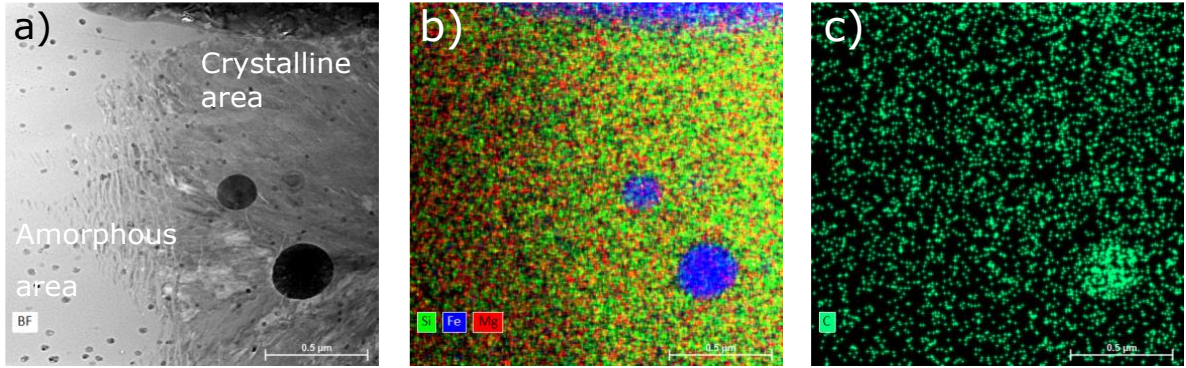

Figure 2: STEM images of quenched silicate melt. a) Bright field image showing two contrasting types of metal inclusions (dark grey) and the presence of both crystalline (Fe-rich bridgmanite) and amorphous regions (the latter were caused by $Ga^+$-ion milling). b) Si, Fe, Mg EDS map showing the homogeneity of the silicate in the crystalline area and the large metal blebs. c) Carbon map of the same area showing homogeneity of C distribution in the silicate, except for the large metal inclusion that has a relatively high C concentration. The edge of the central metal part of the sample (Fig. 1) is seen at the very top of these images.

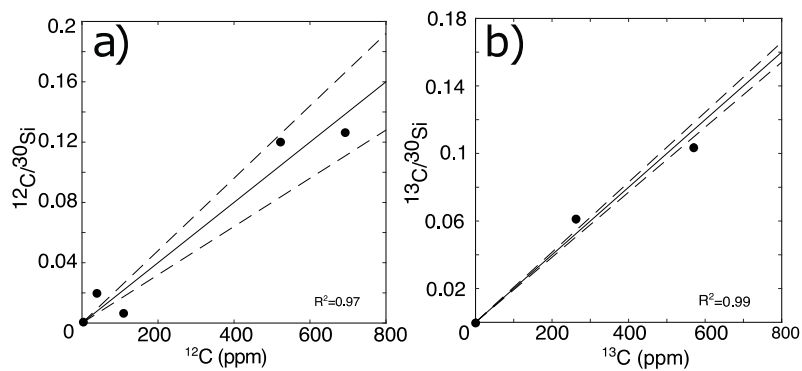

Figure 3: Results from NanoSIMS measurements. a) and b) calibration lines obtained for $^{12}C$ and $^{13}C$ respectively using our standards and two natural samples (d'Orbigny and ALV 981-

R23) and both FTIR and NanoSIMS measurements. Note that for $^{13}$C measurements (b), the two natural samples plot at the origin.

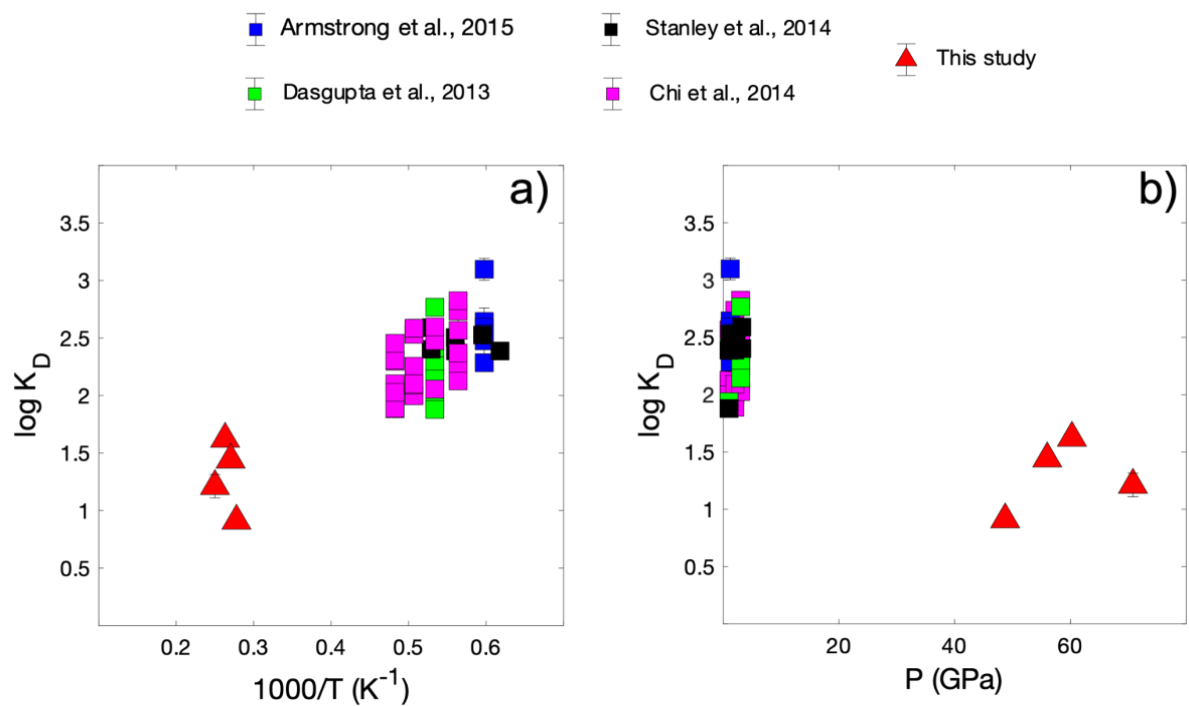

Figure 4: Comparison of our data with results of previous studies with respect to (a) temperature and (b) pressure. The valence of carbon in silicate melt is assumed to be 2+. Uncertainties on temperature and pressure measurements for diamond anvil cell experiments are 300 K and 5 GPa respectively.

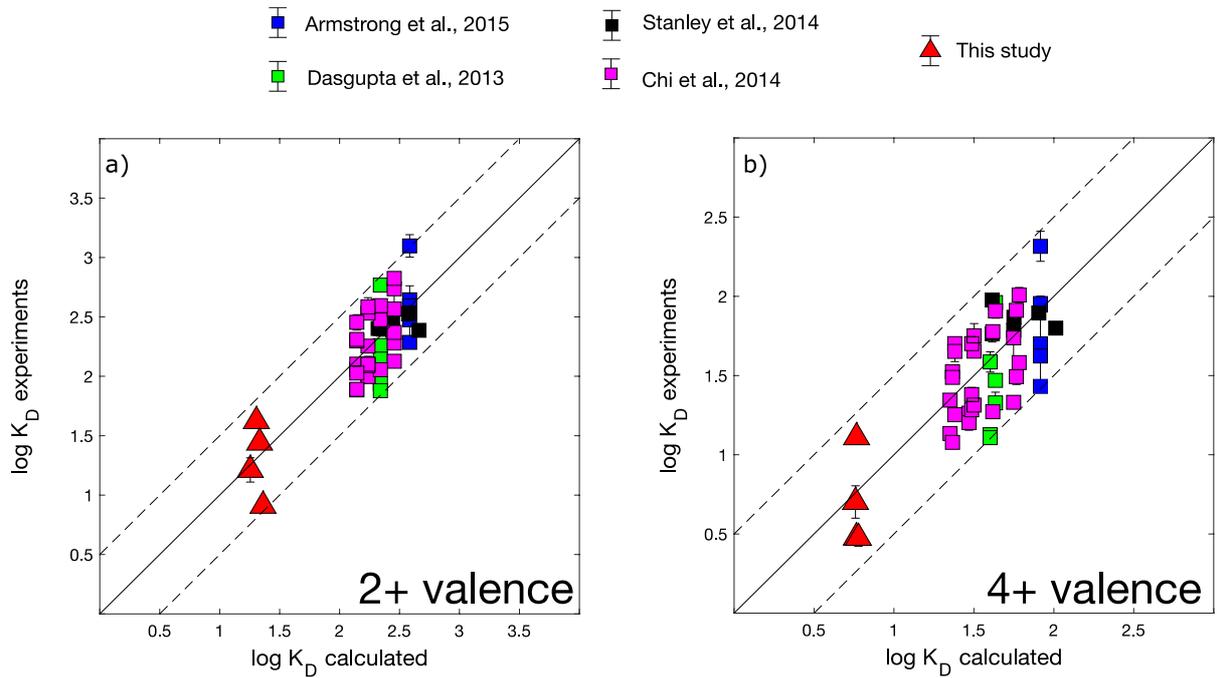

figure 5: Validation of our partitioning model. Comparison of the experimental $K_D$ results (see Eq. 3) and values calculated from (a) Eq. 4 and (b) Eq. 5. In the latter case, two of our data points are superimposed.

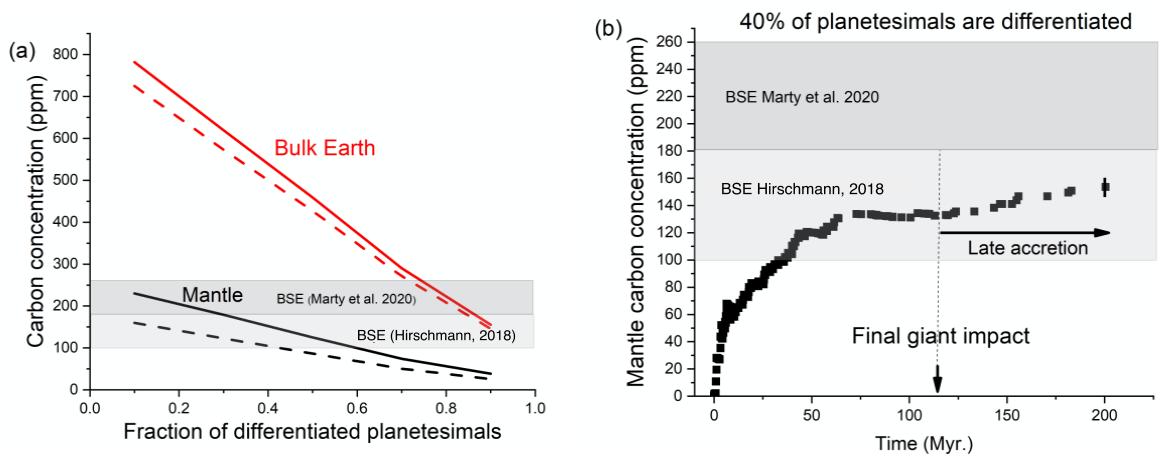

Figure 6: Results of the combined accretion/core-mantle differentiation model. a) Final carbon concentrations in the mantle (black lines) and bulk Earth (red lines) as a function of the fraction

of differentiated planetesimals in the CC and NC regions of the protoplanetary disk. The solid lines show results obtained when undifferentiated CC bodies contain 3.35 wt.% C (CI value) and the dash lines show results when this value is 2.2 wt.% (CM value) (Table 2). The grey regions (in a and b), which strongly overlap, show the BSE concentrations of 140±40 ppm estimated by Hirschmann (2018) and 112-260 ppm estimated by Marty et al. (2020). The concentration of C in Earth's core is 1270±300 ppm when the calculated BSE concentration is 140±40 ppm. b) Carbon concentration in Earth's primitive mantle during accretion as a function of time, where each symbol represents an accretion event. Undifferentiated CC bodies contain 3.35 wt.% C and the fraction of differentiated planetesimals is 0.4. The final uncertainty on the calculated concentration is determined by propagating the uncertainties of the fitted parameters in Eq. 4 and is small because most BSE C is delivered in fully oxidized planetesimals and is largely unaffected by core forming events.

# Supplementary material

**The Metal–Silicate Partitioning of Carbon During Earth's Accretion and its Distribution in the Early Solar System**

I. Blanchard, D.C. Rubie, E. S. Jennings, I. A. Franchi, X. Zhao, S. Petitgirard, N. Miyajima,

S. A. Jacobson, A. Morbidelli

**Laser heating**

The laser beam was focused onto the sample using two NIR-Mitutoyo objective lenses with x20 magnification. Temperature was measured with a 2500i spectrometer and pixis 400 CCD camera from Princeton-Instrument©, from the light collected through both objective lenses. During heating, the thermal radiation was collected from both sides at the same time on the CCD camera via two pinholes at the entrance of the spectrometer. The pinholes, made with a FIB, were 60 microns in diameter and correspond to an area of about 4 microns on the sample due to the x16 magnification obtained from both optical paths. The heating procedure was as follows: 1) the two images from both sides were adjusted so that the two pinholes were on top of each other (or were positioned on the same object). 2) The lasers from both sides were aligned on each side on top of each other and with respect to the pinholes. 3) The power was adjusted independently for each laser, so the temperature difference from both sides was limited to few tens of kelvins. During this stage the pinhole image position could be adjusted to stay on the hottest spot of laser heating. 4) Rapid quenching was achieved by switching off the lasers.

**EPMA analysis**

Major element concentrations in both the quenched metal and silicate phases were analyzed by EPMA. We calibrated using wollastonite, spinel and olivine for Ca, Al and Si, Mg, Fe in the silicate with a current of 15 keV and 15 nA, a focused beam and a phi-rho-Z correction. For the metallic phase, we used a pure Fe wire, FeSi, $Fe_2O_3$ and $Fe_3C$ standards to calibrate for Fe, Si, O and C respectively, at 15 keV with a 25 nA beam current, again using a focused beam and phi-rho-Z correction. The counting time was 10 seconds on the background, and 20 seconds on the peak for all elements, and the samples and the $Fe_3C$ standard were not carbon coated. Analyses performed on the quenched silicate melt highlighted high concentrations of FeO (14 to 27 wt.%), as observed previously in similar studies (Blanchard et al., 2017; Siebert et al., 2012). In the metallic phase, we observed a range of carbon concentrations, from ~5 to 9.5 wt.%, depending on the sample. In addition, the metallic phase contains 2.8-7.5 wt.% oxygen and 0.3-5.6 wt.% silicon.

Analyzing carbon using the microprobe requires special care. In order to quantify the carbon content of both the experimental metals and the metal starting material, we synthesized a $Fe_3C$ carbide primary standard (Jennings et al., 2021). The carbide was fixed to a metallic stage using silver paint and copper tape, and the LH-DAC samples were not coated. During C analysis, we took care by repeating the measurements on the metallic standards to quantify background levels (i.e. possible contamination) inside the EPMA. We followed a similar method for quantifying carbon to that of Dasgupta and Walker (2008), with tests and procedures detailed by Jennings et al. (2021). We performed repeated measurements of pure iron wire to check for carbon contamination, and observed a background for carbon of 0.49±0.07 wt.% which is well below the concentration of carbon present in the metallic phase of our samples. The same was done with pure FeSi, for which we measured 0.36±0.05 wt.% of carbon in the standard. To check for oxygen contamination, we measured a background of 0.04±0.04 wt.% in the Fe wire and 0.15±0.05 wt.% in FeSi. These values are again well below the concentrations measured in our samples. Standard analyses were consistent and stable over the whole duration of the analytical sessions.

**NanoSIMS measurements**

The mass resolving power (m/Δm) for both spot and imaging mode was set to 9000 (CAMECA definition), sufficient to resolve all interferences from neighbouring mass peaks, such as $^{13}$C from $^{12}$CH.

Differences in the geometry and charge compensation between the samples and standards has the potential to lead to unwanted effects in the NanoSIMS measurements. Variations in the secondary ion yield of $^{30}$Si were greater than may have been anticipated from chemical variation, indicating variation in the efficiency in generating secondary ions, which could have the potential to modify measured C/Si ratio. In order to test this, we performed two different tests on the basaltic glass standard (ALV981-R23), one adjusting the electron gun (e-gun) and the other adjusting the Z distance (focus) in order to provide a simulation of deviations/distortion in the electrical field around the sample site. The tests were conducted with the same probe and setup as used for the sample analysis. For the e-gun tests, the e-gun emission was adjusted from the optimised emission current of 0.15 mA down to 0.05 mA, which resulted in the $^{30}$Si count rates decreasing from 42 kcps to ~8 kcps, however the $^{13}$C/$^{30}$Si (and $^{12}$C/$^{30}$Si) ratios were quite consistent. For the Z tests, we changed Z from 2800 to 3300: $^{30}$Si only increased from 40 kcps to 51 kcps. Again, the $^{13}$C/$^{30}$Si (and $^{12}$C/$^{30}$Si) ratios were unchanged, with the 2SD about 2.5% and 10% respectively. We present those results in figure S1. This figure highlights that there is no evidence for any change in the measured C/Si ratio within the uncertainty of the measurements, with the 2SD level at about 8% of measured ratio (primarily reflecting the internal precision on individual measurements). Hence, there is good confidence in the measured C/Si ratios, particularly with the modified sample mounts, with a conductive surface immediately behind the FIB slice.

We report in Table S1 the composition of the starting materials and the standards that were used in this study. The concentrations of $^{13}$C and $^{12}$C in B1145 and B1147 were calculated by first obtaining the bulk C concentrations of the standards using FTIR, then by using the $^{13}$C and $^{12}$C counts at the NanoSIMS we could estimate the concentrations of the respective isotopes. For the D'Orbigny glass and ALV 981-R23, we used the values proposed by Hekinian and Walker (1987), Macpherson et al. (1999) and Varela et al. (2003) and used the natural abundances of $^{13}$C and $^{12}$C (1.1 and 98.9% respectively).

We have also used a San Carlos olivine which contains less than 1 ppm C in order to measure the background. We could determine that the background is approximately 5 ppm $^{12}$C and 0.05 ppm $^{13}$C, which makes any consideration of the background irrelevant for our measurements. In order to assess the uncertainties on our measurements, we have calculated the uncertainty associated with our calibration curve, and propagated these uncertainties into the calculated $^{12}$C and $^{13}$C abundances.

We acknowledge that, ideally, the concentration of carbon in the standards and in the samples should not be too different. Sample BAS C 39 contains 2805 ppm of carbon, which is higher than the maximum value that we measured in our standard that contains the most carbon (1263 ppm of carbon in B1147). This results in higher uncertainties on both $^{13}$C and $^{12}$C measurements for this sample (see Table 1). We note that sample BAS C 39 is the sample containing the highest carbon content in the silicate and in the metal. Since the carbon content of the silicate shows high abundances in both $^{13}$C and $^{12}$C, this is probably due to the small heterogeneity of the starting material, as shown in Table S1.

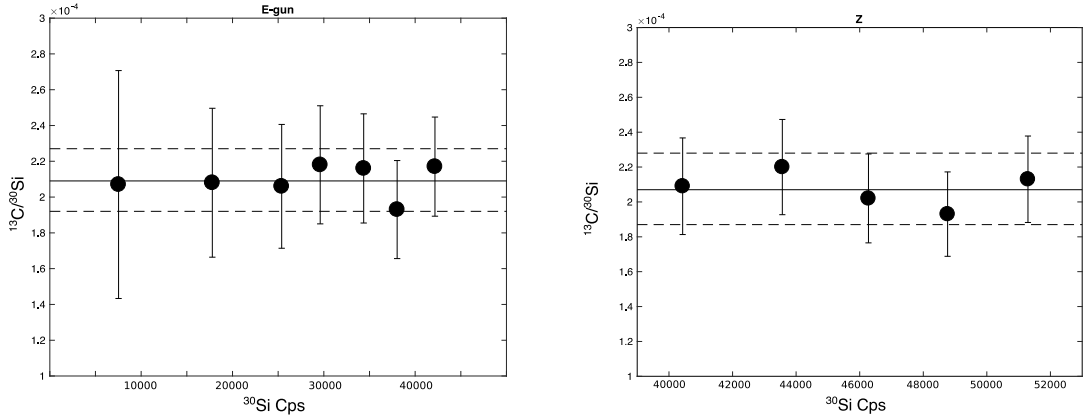

Figure S1: Illustration of the absence of the effects of changing the electron gun from the optimised emission current of 0.15 mA down to 0.05 mA (left) and the focus from 2800 to 3300 (right) at the NanoSIMS on the measurements of $^{13}$C/$^{30}$Si.

**Evaluating chemical equilibrium**

In order to evaluate if chemical equilibrium was achieved in our experiments, we compared the equilibrium constant $K$ for silicon ($K_{Si}$) of our samples with results from the literature. We used $K_{Si}$ because it is known to have a strong temperature dependence and small pressure and compositional dependencies (Mann et al., 2009; Siebert et al., 2012). The values of $K_{Si}$ are $K_{DSi}$ corrected for metal activities using the MetalAct website (http://www.earth.ox.ac.uk/~expet/metalact/, Wade and Wood, 2005) which is based on the formalism of Ma (2001). We compare our data with results obtained over a broad range of temperatures (from 1873 to 4500 K) in Fig. S2. Our results and those obtained previously show a good correspondence which demonstrates that our samples closely (or fully) achieved chemical equilibrium with respect to silicon partitioning during the experiments. This means that equilibrium carbon partitioning was also achieved because carbon diffusion in liquid metal is about 2.5 times faster than silicon diffusion (Rebaza et al., 2021).

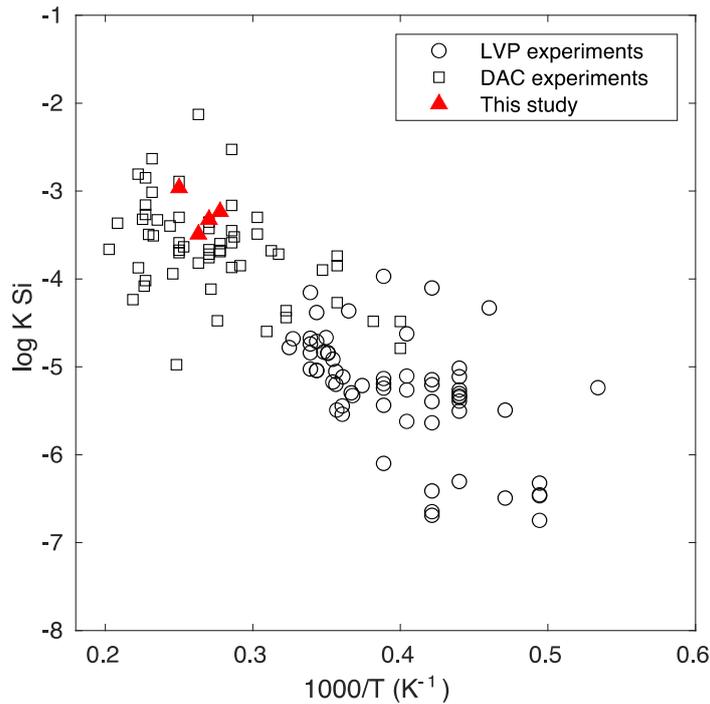

Figure S2: Si-Fe equilibrium constant as a function of inverse temperature for Large Volume Press experiments (LVP, data from Ito et al., 1995; Gessmann C. and Rubie, 1998; Corgne et al., 2008; Mann et al., 2009; Siebert et al., 2011; Siebert et al., 2012; Tsuno et al., 2013; Fischer et al., 2015), Diamond Anvil Cell experiments (DAC, data from Badro et al., 2018; Blanchard et al., 2017; Bouhifd and Jephcoat, 2011; Chidester et al., 2017; Du et al., 2017; Jackson et al., 2018 and this study.

**Comparison with the results of Fischer et al. (2020)**

Here, we compare the results of the current study with those of the recent study of Fischer et al. (2020) in which the metal–silicate partitioning of carbon was studied at pressure and temperature conditions similar to those of the present study. Despite similar $P-T$ conditions, our results are very different, and we think that this is strongly related to the way carbon was measured in the Fischer et al. (2020) study. As discussed in the main text, Fischer et al. (2020) used two methods to measure the concentration of carbon in the quenched silicate melt of their samples: electron microprobe (EPMA) and NanoSIMS. The two methods resulted in very different carbon concentrations with higher concentrations measured with NanoSIMS than with EPMA.

As stated in the main text, we did not attempt to measure the concentration of carbon in the silicate phase of our samples by EPMA, for the following reasons.

1) There are no suitable standards available. Fischer et al. (2020) used a carbide ($Fe_3C$) standard to measure carbon in the silicate phase. Considering lack of similarities between the matrix of the standard and that of the sample, we believe that this could generate problems when quantifying C in the silicate phase.

2) We estimated the background level for carbon measurements using EPMA and found a minimum of 0.36±0.05 wt.% (see in the EPMA analysis section above). This value is higher than the maximum value measured in our samples (sample BAS C 39 containing 0.28 wt.% C). Hence, it would not have been possible to measure reliably the carbon content of the silicate phase of this sample using EPMA.

As discussed above, it is important to obtain a calibration curve for NanoSIMS that spans the concentrations of the element that is being analyzed. In the case of Fischer et al. (2020), they state that their standards contained 45 to 1500 ppm of C, whereas they determined concentrations of 2.3 to 6.6 wt.% carbon in the silicate phase using NanoSIMS. The discrepancy between the concentration of C in their standards and in their samples can also be a substantial source of error in their measurements.

Another source of discrepancy between the two studies can be linked to different starting material, since Fischer et al. (2020) used olivine whereas we used a synthetic basalt.

We compare in Fig. S3 the values of log $K_D$ determined in this study and with the two different sets of log $K_D$ values determined by Fischer et al. (2020). The values determined by Fischer et al. plot well below the ones determined in the current study and the $D_C$ values they obtained by NanoSIMS are especially low. and differ from our results by almost two orders of magnitude. The reason for the large differences, as discussed in the main text, is that they analyzed the entire silicate melt region including the large metallic blebs, as can be seen in their Figure 1. In our study, we avoided including the large blebs of metal that are also present in all our experiments performed at similar $P-T$ conditions.

We also compare in Fig. S4 the experimental values of $D_C$ obtained in our study with values predicted by the two models of Fischer et al. (2020) based on our experimental conditions and phase compositions. The first model is based on their NanoSIMS measurements together with low-pressure data from the literature, whereas the second is derived from their EPMA measurements together with the same low-pressure data from the literature. Partition coefficients predicted by the model based on NanoSIMS measurements (Fischer et al., 2020, Eq. 1) are about two orders of magnitude lower than our experimental values. Partition coefficients predicted by the model based on EPMA measurements (Fischer et al., 2020, Eq. 2) are about one order of magnitude lower than our experimental values.

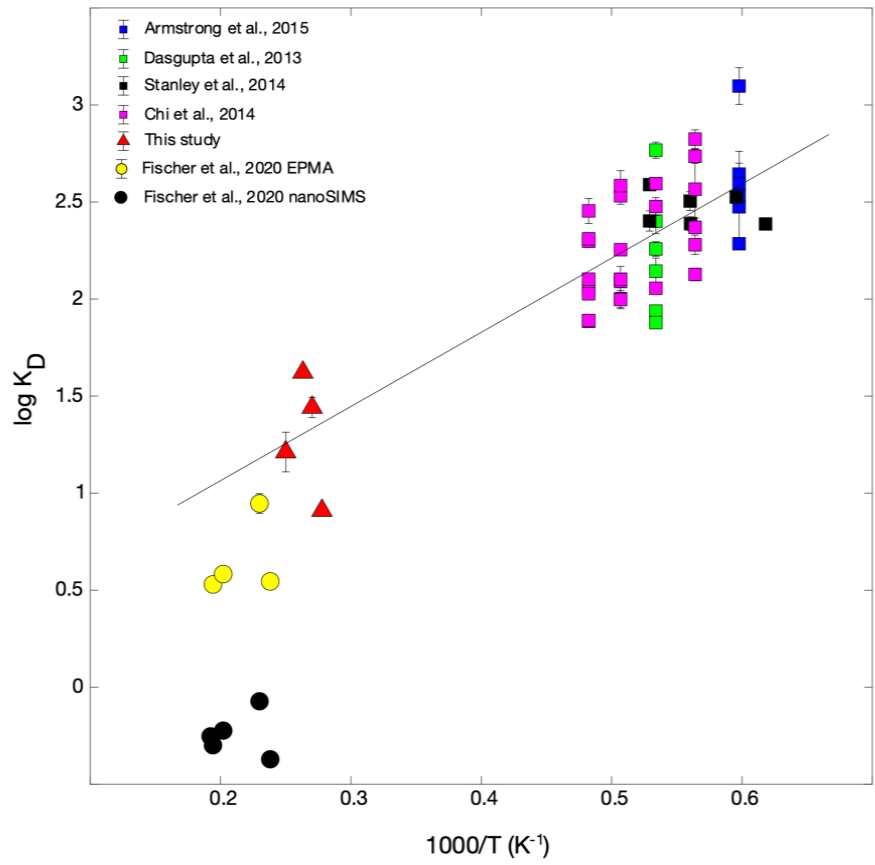

Figure S3: Same data as presented in Fig. 4 of the main paper, but here including the results of Fischer et al. (2020) measured by NanoSIMS and EPMA. $K_D$ is calculated as $D_C/D_{Fe}$ (see main text). The regression we obtained in this study for a valence of 2+ (Eq. 4 in the main text) is shown as a straight line.

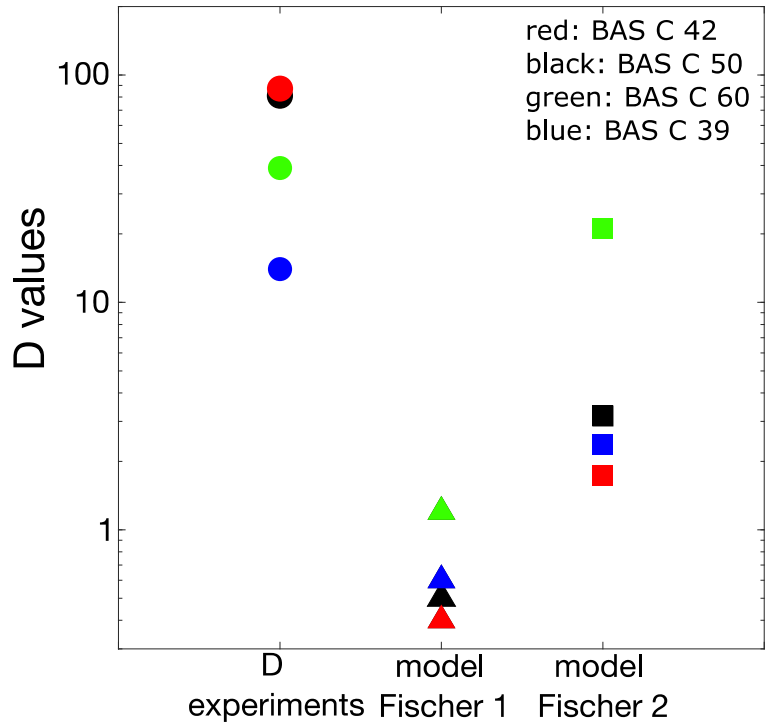

Figure S4: Comparison of the *D* values (in wt%) obtained from our experimental dataset with predicted D values calculated for the same conditions of *P-T*, $fO_2$ and metal composition using equations 1 and 2 from Fischer et al. (2020). Each sample is represented by the following colors: red for BAS C 42, black for BAS C 50, green for BAS C 60, blue for BAS C 39.

**Continuous core formation model**

We have formulated a simple continuous core formation model that has been applied in many geochemical studies (e.g. Wade and Wood, 2005; Wood et al., 2006; Righter et al., 2018b). Earth is assumed to accrete material in small (1%) increments and each batch of accreted metal is assumed to equilibrate chemically with the entire mantle. The equilibration pressure is assumed to correspond to conditions at the base of the magma ocean and increases as the planet grows. The temperature at the base of the magma ocean, which is constrained by the peridotite solidus/liquidus, also increases as the Earth grows. For simplicity, we set the $fO_2$ during accretion at IW-2.4 as given by the current mantle/core FeO/Fe ratio (Corgne et al., 2009).

The concentration of carbon in the BSE today is estimated to lie between 100 and 260 ppm (Hirschmann, 2018; McDonough and Sun, 1995; Marty et al., 2020) and the bulk Earth carbon content has been proposed to be 520 (±210) ppm (Marty, 2012) or 730 ppm (McDonough, 2014). Here, we assume that each stepwise-accretion step delivers material with a bulk C content that leads to the final bulk Earth estimate of McDonough (2014). The partition coefficient of carbon is calculated after each equilibration step using Equation 4 in the main text, and the resulting concentrations of carbon present in both mantle and core are thus derived. Carbon is assumed to be present throughout accretion.

The Earth is believed to have been bombarded by bodies comprised of chondritic material after core formation ceased (the so-called late accretion stage which resulted in the addition of the "late veneer") in order to explain the concentrations of highly siderophile elements present in the mantle (Chou et al., 1983; Kimura et al., 1974; Turekian and Clark, 1969). The amount of material that was delivered to Earth during late accretion was about 0.6% of Earth's mass according to Walker (2009), and was not incorporated into the core, but remained in the mantle. Here, we took the value of 0.6% of Earth's mass because this is typically the value obtained by considering that the HSEs were only added during late accretion (the value of 0.3% that is specified in the main text resulted directly from the N-body simulation).

Chondrites are carbon-rich, with up to 3.35 wt.% of carbon in CI chondrites (Wasson and Kallemeyn, 1988), so their delivery to Earth brought a substantial amount of carbon. We incorporated the effect of such late accretion in our model by assuming that the carbon content of the late veneer matches the approximate mean carbonaceous chondrite (CI, CM, CO and CV) value of 1.6 wt.%.

Results of the continuous core formation model are presented in Fig. S5. The combined effects of core formation and the subsequent addition of the late veneer are needed in this model in order to reproduce the carbon content of the BSE. Core–mantle differentiation effectively strips the mantle of all carbon and the addition of C during late accretion is required to achieve the BSE carbon concentration. This result is based on the assumption of a constant $fO_2$ of IW-2.4 during accretion whereas successful core formation models have generally concluded that $fO_2$ was 2-3 orders of magnitude more reducing during the initial 40-60% of accretion (e.g. Wade and Wood, 2005; Rubie et al., 2011). The effect of more reducing conditions would be to decrease even more the mantle concentration of C prior to the late accretion stage.

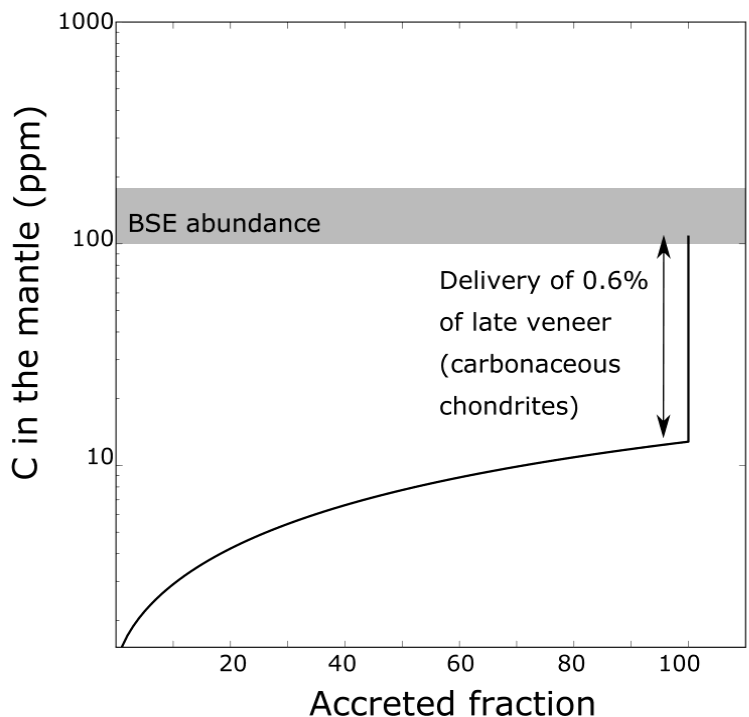

Figure S5: Evolution of the concentration of carbon in the mantle during Earth's core–mantle differentiation and the late accretion stage based on a continuous core formation model. An estimated BSE concentration of 100-180 ppm (Hirschmann, 2018) is shown.

**Accretion/core formation model - sources of uncertainties.**

1) Our parameterizations were obtained using a basaltic silicate melt composition with (low pressure) nbo/t values in the range of 0.56-1.34 (Table 1) which are lower than that of a peridotitic melt (2.5) in a magma ocean. Assuming the dependence of $D$ on nbo/t determined by Fischer et al. (2020), a nbo/t value of 2.5 would increase the final mantle C concentrations of Fig. 6 by only 6-10 ppm. This extrapolation is uncertain because nbo/t only describes melt structure at very low pressures and does not take account of pressure-induced structural changes (Mysen and Richet, 2005). A basaltic melt is expected to become increasing depolymerized with increasing pressure so the structural differences between basaltic and peridotite liquids at the high pressures of our experiments may be small.

2) The valence of C in the silicate melts of our study is uncertain and may be either 2+ or 4+; partition coefficient expressions have been determined for both cases (Eqs. 4 and 5). The results of Fig. 6 have been obtained assuming 2+ but if a valence of 4+ is assumed, the final mantle concentrations change by only a few ppm. This is because almost all BSE carbon is delivered in fully oxidized material that is not affected by core formation, as discussed in the main text.

3) As stated in the main text, there is no model that describes quantitatively the chemical consequences when undifferentiated metal-bearing planetesimals from the NC region are accreted. The small metal grains in such material will be more widely dispersed in the magma ocean and may equilibrate with a larger fraction of the mantle compared with predictions of the Deguen et al. (2011) model. The results presented in Fig. 6 have been obtained using the Deguen et al. (2011) model (based on the mass fraction of metal in accreting undifferentiated bodies). If the metal equilibrates with a fraction of the mantle that is significantly larger than that predicted by the Deguen et al. model (e.g. larger by a factor of 5-10), final mantle C concentrations are reduced by up to a few tens of ppm (Fig. S6). Thus, the approach we have taken provides upper limiting values of final mantle C concentrations (Fig. S6).

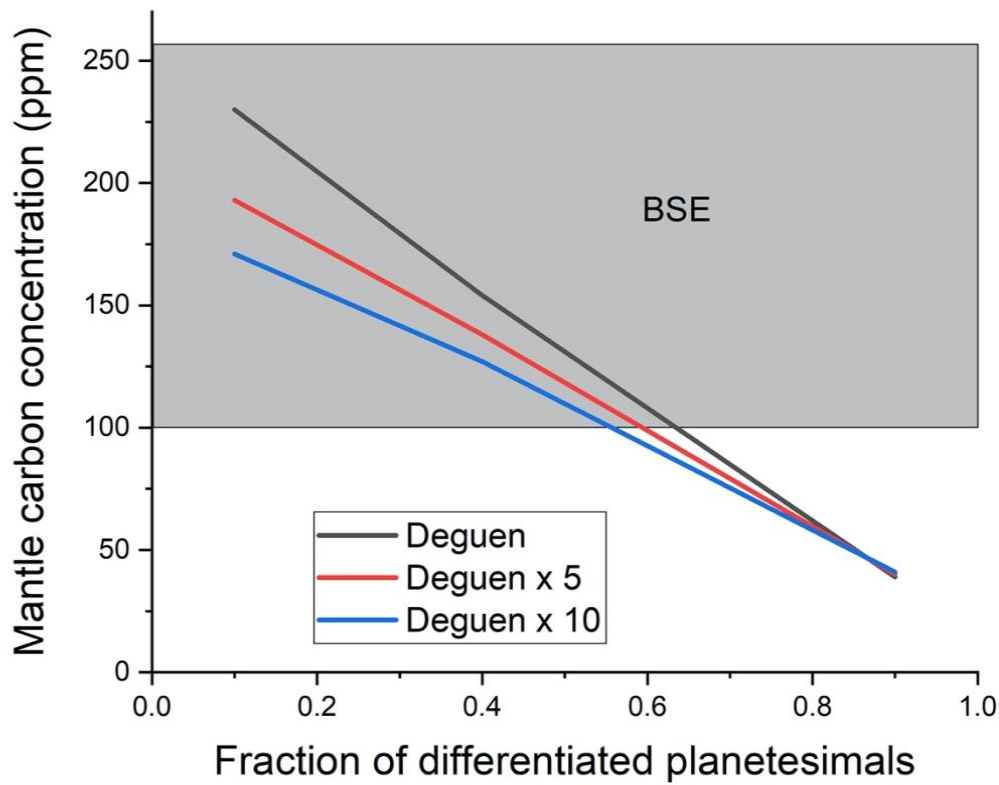

Figure S6: Final carbon concentration of Earth's mantle as a function of the fraction of planetesimals that are differentiated. The line "Deguen" shows the results (as also shown in Fig. 6a) when the metal of undifferentiated planetesimals equilibrates with a fraction of Earth's mantle as predicted by the model of Deguen et al. (2011). The lines "Deguen x 5" and "Deguen x 10" show the results when the fraction of equilibrated mantle that is predicted by Deguen et al. is increased by factors of 5 and 10 respectively.

4) Partition coefficients predicted by Eqs. 1 and 2 of Fischer et al. (2020) are generally lower than those predicted by our parameterizations by 1-2 orders of magnitude (Figs. S3 and S4). In addition, the oxygen fugacity coefficients of -0.211(92) and -0.238(94) determined by Fisher et al. are very low and would indicate that carbon dissolved in silicate liquid has a valence of 1+ whereas a valence of 2+ is much more likely (Armstrong et al., 2015; Grewal et al., 2020; Malavergne et al., 2019; Yoshioka et al., 2019).

The result of our accretion/differentiation model using the parameterization of Eq. 1 in Fischer et al. (2020) is shown in Fig. S7. Because of the low partition coefficients, the mantle C concentration reaches 400 ppm by the end of accretion. Oxygen fugacities are often low during the early stages of accretion (e.g. $\Delta$IW-4) because of the reduced state of accreting bodies. The initial steep increase in the C concentration is caused by the low $fO_2$ coefficient which has a large effect in lowering the partition coefficient when conditions are reduced.

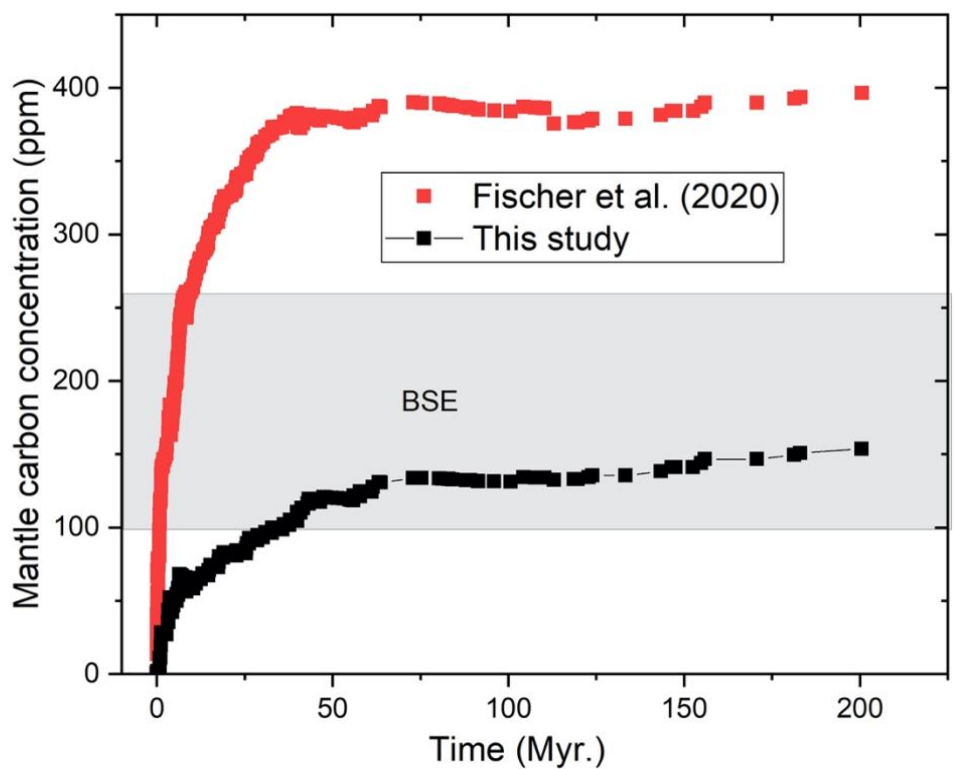

Figure S7. Mantle carbon concentration in Earth's mantle during accretion as a function of time where each symbol represents an accretion event. As in Fig. 6b, CC undifferentiated

bodies contain 3.35 wt.% C and the fraction of differentiated planetesimals is 0.4. Results are shown with metal–silicate partition coefficients defined by Eq. 1 in Fischer et al. (2020) and Eq. 4 of the present study respectively.

**Supplementary Table**

Table S1: Composition of the synthetic basaltic starting silicate used for the LH-DAC experiments ("Basalt"), the standards for the NanoSIMS synthesized with piston-cylinder at 2 GPa and 1600°C ("B1145" and "B1147"), the carbon-free starting material for piston-cylinder experiments ("B1") and the starting metal used in LH-DAC experiments. Composition in major elements were obtained by microprobe measurements. We also report the composition of the two natural samples that were used at the NanoSIMS as standards. b.d. stands for below detection limit.

| Samples | Basalt | B1145 | B1147 | B1 | D'Orbigny glass[a] | ALV 981-R23[c] | Metal | |
|---|---|---|---|---|---|---|---|---|
| *wt.%* | *N=10* | *N=7* | *N=7* | *N=7* | - | - | | *N=14* |
| $SiO_2$ | 52.18 (0.15) | 48.05 (0.18) | 50.89 (0.27) | 53.56 (0.13) | 41.7 | 49.63 | C | 5.67 (1) |
| $Na_2O$ | 1.90 (0.03) | 1.80 (0.06) | 1.86 (0.06) | 2.03 (0.03) | - | 2.91 | Fe | 91.61 (1.31) |
| CaO | 10.92 (0.07) | 9.47 (0.10) | 10.03 (0.08) | 10.70 (0.05) | 23.3 | 11.61 | | |
| FeO | 11.35 (0.04) | 13.65 (0.20) | 10.71 (0.06) | 10.04 (0.06) | 11.0 | 8.16 | | |
| $Al_2O_3$ | 14.66 (0.07) | 16.37 (0.11) | 15.22 (0.10) | 15.48 (0.05) | 19.6 | 16.60 | | |
| MgO | 7.45 (0.07) | 7.22 (0.08) | 7.57 (0.10) | 7.73 (0.06) | 1.87 | 8.36 | | |
| $^{12}C$ (ppm)[e] | - | 522 (78) | 693 (103) | b.d.* | 34 (5.1) | 110 (16) | | |
| $^{13}C$ (ppm)[e] | - | 263 (39) | 570 (85) | b.d.* | 0.36 (0.05) | 1.22 (0.2) | | |
| Total | 98.46 (0.20) | 96.66 (0.19) | 96.41 (0.33) | 99.55 (0.15) | 99.3[b] | 98.96[d] | | 97.28 (0.76) |

*Measured using FTIR (see main text for details)
[a] Composition from Varela et al., 2003
[b] Total includes also 2 wt.% of $TiO_2$, $Cr_2O_3$, MnO
[c] Composition from Hekinian and Walker, 1987; Macpherson et al., 1999.
[d] Total includes $TiO_2$, MnO, $K_2O$ and $P_2O_5$
[e] See text for details on how to obtain $^{12}C$ and $^{13}C$ of standards